# Novel path towards compact laser ion accelerators for hadron therapy: Tenfold energy increase in laser-driven multi-MeV ion generation using a gas target mixed with submicron clusters

Y. Fukuda[1], A. Ya. Faenov[1,2], M. Tampo[1], T. A. Pikuz[1,2], T. Nakamura[1], M. Kando[1], Y. Hayashi[1], A. Yogo[1], H. Sakaki[1], T. Kameshima[1], A. S. Pirozhkov[1], K. Ogura[1], M. Mori[1], T. Zh. Esirkepov[1], A. S. Boldarev[3], V. A. Gasilov[3], A. I. Magunov[4], R. Kodama[5], P. R. Bolton[1], Y. Kato[1,6], T. Tajima[1,7,8], H. Daido[1], and S. V. Bulanov[1,4]

[1]Kansai Photon Science Institute and Photo-Medical Research Center,
  Japan Atomic Energy Agency, Kyoto, 615-0215 Japan.
[2]Joint Institute of High Temperatures, Russian Academy of Sciences,
  Moscow, 125412 Russia.
[3]M.V. Keldysh Institute of Mathematical Modeling, Russian Academy of Sciences,
  Moscow, 125047 Russia.
[4]A.M. Prokhorov Institute of General Physics, Russian Academy of Sciences,
  Moscow, 117942 Russia.
[5]Faculty of Engineering and Institute of Laser Engineering, Osaka University,
  Osaka, 565-0871 Japan.
[6]The Graduate School for the Creation of New Photonics Industries,
  Shizuoka, 431-1202 Japan.
[7]Sektion Physik, Ludwig-Maximilians-Universitaet Muenchen,
  D-85748 Carching, Germany.
[8]Max-Planck-Institut fuer Quantenoptik, D-85748 Garching, Germany.

**Abstract:**
  We demonstrate generation of 10-20 MeV/$u$ ions with a compact 4 TW laser using a gas target mixed with submicron clusters, corresponding to tenfold increase in the ion energies compared to previous experiments with solid targets. It is inferred that the high energy ions are generated due to formation of a strong dipole vortex structure. The demonstrated method has a potential to construct compact and high repetition rate ion sources for hadron therapy and other applications.

Recent development of ultrashort-pulse, high peak power laser systems enables us to investigate high field science under extreme conditions (1). Ion acceleration with intense laser pulses has been one of the most active areas of research in high field science during the last several years (2, 3), because it has a broad range of applications including cancer therapy (4, 5), isotope preparation for medical applications (6), proton radiography (7), and controlled thermonuclear fusion (8).

Multi-MeV ions have been generated in overdense plasmas with thin foil targets (9-11), where proton acceleration has been interpreted as target normal sheath acceleration (TNSA) (12) (see for details Ref. (13)). Also, generation of multi-MeV ions in underdense plasmas has been demonstrated, where the lower-density plasmas were produced by direct ionization of gas targets (14-16) or by ionization and evaporation of thin solid targets by the laser prepulse (17). In the case of the underdense plasma, it has been shown that a quasi-static magnetic field produced by the fast electron current at the rear plasma-vacuum interface builds up a long-lasting electrostatic sheath which can accelerate and collimate ions (18, 19). Furthermore, monoenergetic ions have been generated using microstructured targets (5, 20-22). The divergence and the energy spread of ions have been controlled with a superconducting magnet (23) or a plasma microlens (24). These advances establish a firmer basis for realizing various applications of laser-driven ions. However significant enhancement of the ion energies is necessary for application to cancer therapy, because the treatment requires proton energies of ~200 MeV.

We describe here a new approach where high energy ions are generated from the irradiation of a gas mixed with submicron-size clusters by a compact ultrashort pulse laser. With this approach, we have succeeded in generation of laser-driven ions of energies approximately tenfold higher than those obtained in previous experiments with solid targets driven by small energy laser systems (25, 26). The cluster target, which consists of solid-density clusters embedded in a background gas, has shown unique properties such as efficient plasma waveguide formation (27) and generation of copious high energy electrons (28). These properties are particularly advantageous for effective acceleration of ions, because the plasma waveguide leads to transport of the intense laser pulse over a long distance and the high electron current creates a strong quasi-static magnetic field (19). Furthermore the cluster target enables high repetition rate generation of high energy ions in the absence of the plasma debris that is typical of solid targets.

The experiment has been conducted using the JLITE-X 4-TW Ti:sapphire laser at JAEA-KPSI. A schematic of the experimental set up is shown in Fig. 1A. The laser delivers 40-fs duration (FWHM) pulses of 150 mJ energy at a 1 Hz repletion rate with a temporal contrast near $10^{-6}$. A pulsed solenoid valve connected to a specially designed circular nozzle having a three-stage conical structure with an orifice diameter of 2 mm was used to produce submicron-sized $CO_2$ clusters embedded in a He gas. With the aid of a numerical model (29), the gas parameters were optimized for the production of submicron-sized clusters for a 60-bar gas of 90 % He and 10 % $CO_2$.

The laser pulse was divided into the main pulse and a lower energy probe pulse. The main laser pulse was focused to a spot of 30 μm diameter ($1/e^2$ intensity) with an off-axis parabola of effective focal length, 646 mm. This yields a peak vacuum intensity of $7 \times 10^{17}$ W/cm$^2$. Soft x-ray spectra were acquired using a focusing spectrometer with two-dimensional spatial resolution (30) equipped with a spherically bent mica crystal and a back illuminated CCD camera. The position and timing of the nozzle emission with respect to the laser pulse arrival were adjusted to maximize the intensities of the $He_\beta$ (665.7 eV) and $Ly_\alpha$ (653.7 eV) lines of oxygen (see Fig. 1B). The high energy ions were generated when the laser beam was focused near the rear side of the gas jet and 1.5 mm above the nozzle orifice (see Fig. 2A). At this position, we estimate that solid-density $CO_2$ clusters with an average diameter of 0.37 μm are embedded in the He gas of density $1.5 \times 10^{19}$ cm$^{-3}$ which is about 2 % of the critical density, $n_c$ ($n_c = m_e \omega^2 / 4\pi e^2$). Here we obtain a high cluster density ($3 \times 10^9$ cm$^{-3}$) and a small inter-cluster distance (5 μm). It is expected that significant enhancement of the laser intensity will occur during propagation through the gas, because the peak power of the main laser pulse is well above the critical power for relativistic self-focusing. The shadowgraph image shown in Fig. 2A reveals the formation of a channel of approximately 5 mm in length, substantially longer than the nozzle orifice diameter (2 mm) and the Rayleigh length (900 μm).

The high energy ions were measured with a stack of solid state nuclear track detectors (SSNTD) placed on the laser propagation axis at a distance of 200 mm from the laser focal plane. The SSNTD stack consists of ten sheets of 10-μm thick polycarbonate film and twelve sheets of 100-μm thick CR39 with an area of 40×40 mm. A single 6-μm Al foil was placed in front of this track detector to protect it from damage induced by the transmitted portion of main laser pulse. Ions were accumulated for about six thousand laser shots. Fig. 1C shows a typical image of the etched pits, which was registered in the 11th layer of CR39, observed with a differential interference microscope. Observations of the pit images through the whole layers of CR39 reveal that these pits penetrate through several successive CR39 layers at exactly the same lateral position, and vanish at some layer which corresponds to the depth of Bragg peak for ions in the CR39 stack. The track images also show that these high energy ions are well collimated with a divergence (full angle) of 3.4° in the forward direction.

The energy range of the ions is determined quantitatively from the extent of the tracks recorded in the CR-39 stack by calculating their stopping ranges. Since the target gas is a mixture of He and $CO_2$, highly charged helium, carbon, and oxygen are the possible accelerated ion candidates. We observe the ion tracks in CR39 up to the 11th layer and none in the 12th layer. This penetration depth corresponds to maximum energies of 10, 17, and 20 MeV per nucleon for helium, carbon, and oxygen ions, respectively. We note that there are at least two different sizes of tracks. It is likely that the smaller ones are from helium ions and the larger ones are from carbon and/or oxygen ions.

In order to confirm that the existence of the clusters in the gas jet is responsible for the self-focusing (long channel formation) and the ion acceleration, a larger laser prepulse with the contrast of $10^{-4}$ was intentionally introduced before the main pulse to completely destroy the clusters prior to the arrival of the main pulse. In this case we confirmed that long-distance self-focusing and ion acceleration were not observed. Furthermore, when the laser was focused onto a target composed only of a 60-bar He gas of density $2\times10^{19}$ cm$^{-3}$ (~$0.02n_c$) using the same nozzle, where clusters are not formed due to the weak van der Waals force between the He atoms, the rear part of the channel structure disappeared (see Fig. 2B) and high energy ions were not observed.

Further evidence for high energy ion production was obtained from the time-of-flight (TOF) measurements obtained with a microchannel plate (MCP) detector placed 930-mm downstream of the gas nozzle along the laser propagation axis in place of the SSNTD stack. An electromagnet of field magnitude ~0.15 T was placed between the gas target and MCP to deflect accelerated electrons (with energies below 20 MeV) from the MCP detector. In addition, three layers of the 13-mm thick Al foils were inserted to block the transmitted portion of the main laser pulse and to reduce the incident X ray radiation. Thus helium, carbon, and oxygen ions with the energies exceeding 1.9, 3.2, and 3.6 MeV per nucleon, respectively, could reach the MCP detector. Figure 3 shows an ion energy spectrum thus obtained. The energy scale of the abscissa was calculated assuming the observed ion signals to be those of carbon. The ordinate is the number of the ions observed over 285 consecutive laser shots. The maximum ion energy is measured to be 18.5±1 MeV per nucleon, which is consistent with the energy observed with the CR39 track detectors. It is estimated that the number of ions with the energy exceeding 3.2 MeV per nucleon is approximately $1.7\times10^3$ sr$^{-1}$ for a single laser pulse.

We note that with the cluster target, ions were accelerated up to approximately 20 MeV per nucleon with a laser pulse energy of only 150 mJ. This corresponds to approximately a tenfold improvement of accelerated ion energy compared to previous experiments, where 1.3-1.5 MeV protons were produced at laser pulse energies of 120-200 mJ in solid targets (25, 26).

Two-dimensional particle-in-cell (PIC) simulations were conducted to better understand the ion acceleration process. The laser pulse irradiates the gas density plasma which is composed of the electrons and the ions with $Z/A=1/2$ such as He$^{2+}$, C$^{6+}$ and O$^{8+}$ with the density profile shown in Fig. 4A by the red line. The laser and plasma parameters correspond to those of the present experiment. The simulation was run for a 100-μm thick plasma slab, because high energy ion generation has been observed when the laser pulse is focused near the rear side of the gas jet. The density was kept constant at $0.1n_c$ for $20 < x < 65$ μm, decreasing linearly to $0.02n_c$ in the region, $65 < x < 80$ μm, and kept constant at $0.02n_c$ for $80 < x < 110$ μm.

The laser pulse undergoes the relativistic self-focusing leading to the plasma channel formation filled with the a quasi-static magnetic field of about 35 MG (Figs. 4A-C). Fast electrons accelerated in the channel form a dipole vortex in the slope plasma region of $70 < x < 90$ (18,19), which is associated with the strong bipolar magnetic field as illustrated in Fig. 4B. The magnetic field presses out the cold plasma electrons forming the low density regions surrounded by dense thin shells. The density profile of Fig. 4A resembles the shadowgraph image observed in the experiment (Fig. 2A). As is well known, the fast electrons produce a quasi-static electric field at the plasma-vacuum interface (Fig. 4C, region $x > 110$), which is associated with the acceleration mechanism known as TNSA. However we find in the present case that a much stronger electric field is generated at the shells wrapping the dipole vortex (Fig. 4C, region $70 < x < 90$). Since the electric field reverses sign, the ions are accelerated in the direction perpendicular to the shell surface to the maximum energy of 8.5 MeV per nucleon (Fig. 4D). The electrons have a quasi-thermal energy spectrum with an effective temperature of 1.9 MeV and a maximum energy of ~20 MeV. The simulation shows that the ions are accelerated along the laser propagation axis in a time dependent electric field generated during the magnetic field annihilation. This process is similar to the ion acceleration in plasma pinch discharges. In addition, accelerated ions directed perpendicular to the shell originate from the Coulomb explosion of the shells when the electrons of the shells are expelled by the inhomogeneous magnetic field pressure. Both processes give comparable ion energies. Our simulations suggest that increasing the areal density, *nl*, of the target and the energy of the laser pulse will result in the increase of the ion energies.

The contribution of the TNSA mechanism acting at the plasma-vacuum interface is estimated to be about 2 MeV/*u*. Thus the presence of the dipole vortex structure is essential for high energy ion generation under the present experimental conditions. Our computer simulations indicate that generation of such a magnetic field requires an optimal slope-step profile and optimal plasma density (~$0.1n_c$ in our case). This level of plasma density cannot be created by the background He gas (~$0.02n_c$), but can be created by addition of the $CO_2$ clusters which are initially at solid density and then expand during laser irradiation. In fact, analysis of soft X-ray spectra ($He_\beta$ and $Ly_\alpha$ lines of oxygen) shown in Fig. 1B determines the plasma electron density to be of order $10^{20}$ cm$^{-3}$ which is about $0.1n_c$. Thus the use of a mixture of He gas and $CO_2$ clusters is crucial for securing the proper plasma density and the slope-step profile.

In conclusion, we have demonstrated efficient generation of high energy ions with energies up to 10-20 MeV per nucleon and with a small divergence (full angle) of 3.4° by irradiating a He gas/$CO_2$ cluster mixture with 40-fs laser pulses of only 150 mJ energy at 1 Hz repetition rate. Favorable comparison between experiment and simulation confirms that these ions are generated at the rear side of the target due to the formation of a strong dipole vortex structure. This demonstrated energetic ion production with a compact laser of modest pulse energy can lead to development of the compact, high repetition rate laser-driven ion sources for hadron therapy and other applications.

This work was supported by the Special Coordination Funds (SCF) for Promoting Science and Technology commissioned by MEXT of Japan. We are grateful to Professor Mitsuyuki Abe for his continuous support for this work.

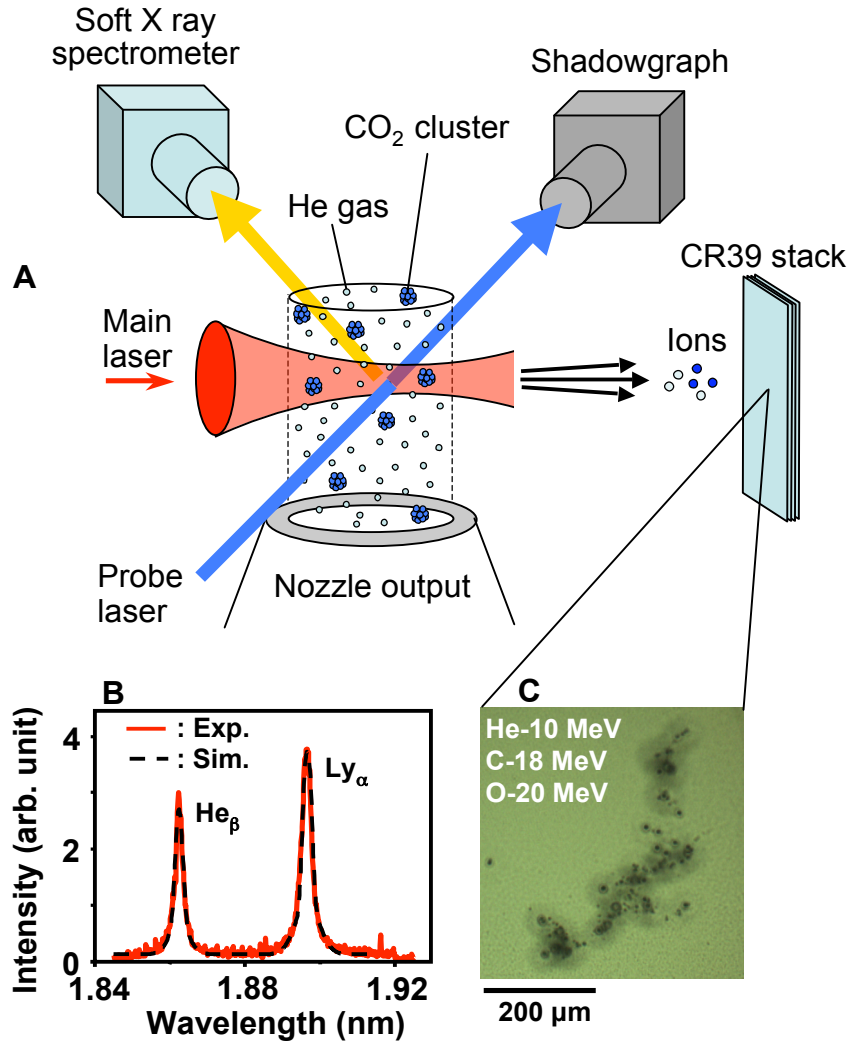

Fig. 1 (**A**) Schematic of the experimental setup. The He gas mixed with submicron $CO_2$ clusters were produced with a nozzle of 2 mm orifice diameter. The main laser pulse was focused to the region near the rear side of the gas jet. The position and timing of the nozzle emission with respect to the laser pulse arrival were adjusted to maximize the intensities of the soft X-ray spectra. The laser propagation in the gas was monitored by the shadowgraph images with the probe laser. The high energy ions were measured with a stack of solid state nuclear track detectors placed on the laser propagation axis. (**B**) Measured soft X-ray spectrum of the $He_\beta$ and $Ly_\alpha$ lines of oxygen (solid curve), and the calculated spectrum (dotted curve) for the near-critical density plasma of order $10^{20}$ cm$^{-3}$ which is about $0.1 n_c$. (**C**) Typical images of the etched pits registered in the 11th layer of CR39, whose penetration depth corresponds to maximum energies of 10, 17, and 20 MeV per nucleon for helium, carbon, and oxygen ions, respectively.

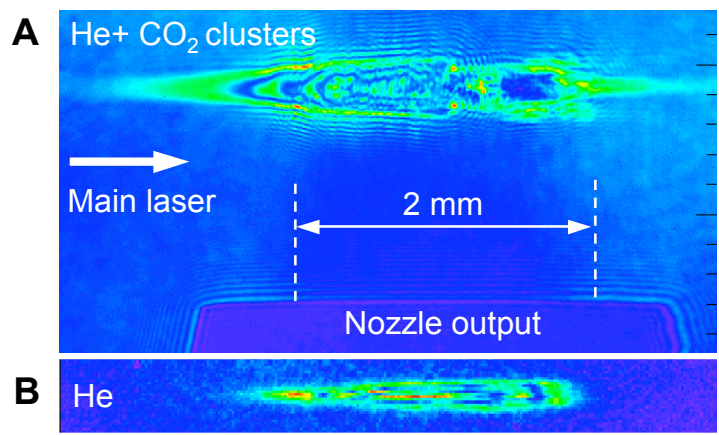

Fig. 2 (**A**) The shadowgraph image for a mixture of He gas and submicron $CO_2$ clusters, revealing formation of a channel of ~5 mm length. (**B**) The shadowgraph image for a target composed only of a 60-bar He gas. The rear part of the channel structure disappeared.

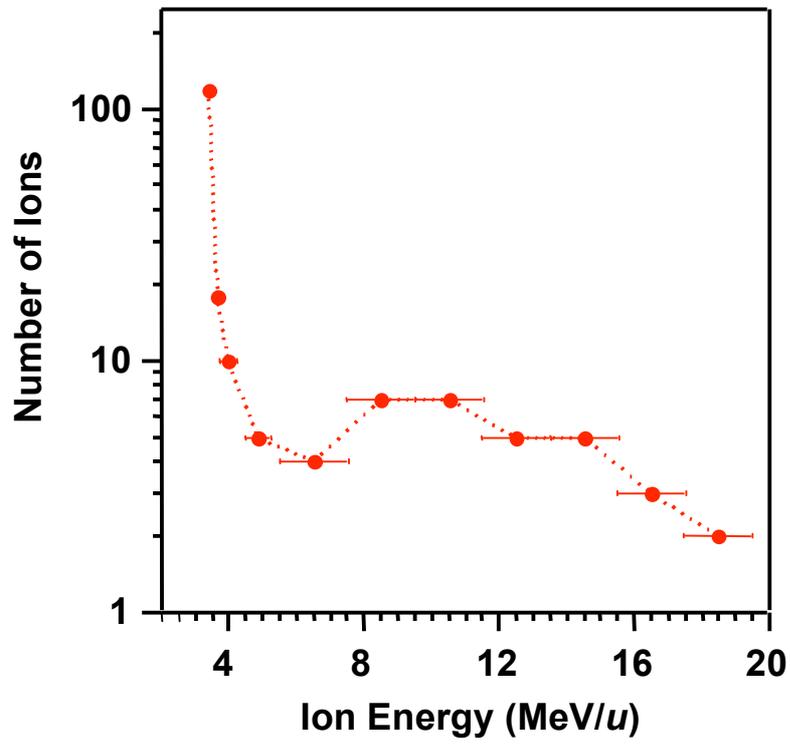

Fig. 3 The ion energy spectrum obtained by TOF. The energy scale of the abscissa was calculated assuming the observed ion signals to be those of carbon. The ordinate is the number of the ions observed over 285 consecutive laser shots.

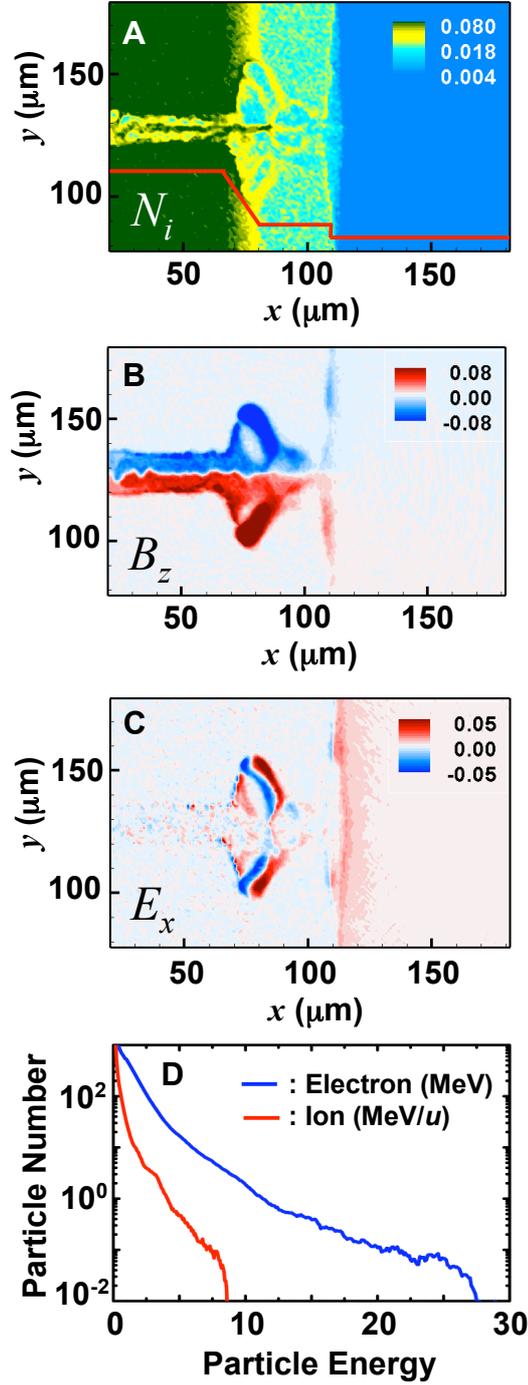

Fig. 4 The results of 2D PIC simulation for a 100-μm thick plasma slab of the density ~$0.1n_c$. The density profile is shown by the red line in (**A**). (**A**): the electron density normalized by $n_c$, (**B**): the magnetic field $B_z$ normalized by the laser field, and (**C**): the longitudinal electric field $E_x$ normalized by the laser field at $t$=900 fs, respectively. (**D**) shows the calculated spectra of the ions in unit of MeV/$u$ and the electrons in MeV at $t$=6 ps.